\documentclass[prb,10pt,twocolumn,amsmath,amssymb]{revtex4-2} 
\usepackage{amsmath}  
\usepackage{amsfonts} 
\usepackage{graphicx} 
\usepackage{hyperref}
\usepackage[utf8]{inputenc} 
\usepackage{amsthm} 
\usepackage{color}

\newcommand{\eq}[1]{{Eq.~(\ref{#1})}} 

\begin{document}
\raggedbottom

\title{Modeling the bounce of a gas-filled ball}

\author{Martin Allen}
\email{mallen.uni@gmail.com}
\affiliation{School of Computer Science, University Of St Andrews,
  College Gate,
  KY16 9AJ, St Andrews, United Kingdom}
\author{Bruce Allen}
\email{bruce.allen@aei.mpg.de}
\affiliation{Max Planck Institute for Gravitational Physics (Albert
  Einstein Institute), Leibniz Universit\"at Hannover, Callinstrasse 38,
  D-30167, Hannover, Germany}

\date{\today}
\begin{abstract}
\noindent
The coefficient of restitution $\epsilon$ characterizes the energy
retained when a ball bounces, and can easily be measured in an ``at
home'' experiment. For thin-walled gas-filled balls such as
basketballs, we construct a simple two parameter model to describe how
$\epsilon$ changes as a function of the ball's internal pressure.  A
comparison with data shows good agreement. With additional
assumptions, the model also predicts how $\epsilon$ changes as a
function of temperature. A comparison with tennis ball data shows
surprisingly good agreement.
\end{abstract}

\maketitle

\section{Introduction}

The bouncing of a ball demonstrates important features of mechanics,
including the conservation of energy and momentum, the conversion of
energy between potential and kinetic, and the loss of mechanical energy 
through conversion to thermal energy and sound.  An interesting
experiment is to drop a ball from a fixed height and to measure how
high it bounces.  The less energy is lost during the bounce, the
closer the ball returns to its starting height.  Accurate measurements
are possible using a mobile phone to video the bounce, and then
examining individual frames.

By making simplifying assumptions, one can construct physical models
to describe such behavior.  While these oversimplify or ignore complex
phenomena, they can still provide a remarkably accurate description.

Here, starting from the principle of energy conservation, we derive a
model to describe the bounce, and compare this to data collected while
varying the pressure inside a basketball.  We then extend the model to
describe how the bounce is affected by temperature.  This is checked
against data collected while varying the temperature of a tennis
ball~\footnote{The first author developed an initial model and
collected the temperature data for his IB Diploma Physics Internal
Assessment.}.

Others~\cite{Bridge1,Bridge2,RoseCoe,Wadhwa,Georgallas} have done such
experiments with a variety of balls.  Often, the data is fit to an
ad-hoc model which is not derived from physical laws, a point also
noted in \cite{Georgallas}.  While our model is an oversimplification,
it nevertheless provides an excellent description of the behavior.

\section{Bouncing Balls}

The Internet offers many slow-motion videos of bouncing balls.  Some
are remarkable, for example, a solid rubber golf ball bouncing off a
steel plate at very high speeds~\cite{GolfBall,GolfBall2}.  Gas-filled
balls such as tennis balls, soccer balls, and basketballs exhibit
similar behavior~\cite{TennisBall}.

These images show that balls are compressed and deformed when they
bounce.  Their kinetic energy is converted into potential energy and
stored: the deformed ball behaves like a compressed spring.  It then
``jumps back'' to a round shape, converting stored potential energy
back into kinetic energy, and leaping into the air. In this process,
some energy is converted into thermal energy and sound, and consequently each bounce of
the ball is lower than the previous one.  A detailed discussion of the
mechanisms at work, and citations to the literature, can be found in a
highly cited paper by Cross~\cite{cross} and work by
Bridge~\cite{Bridge1,Bridge2}.

How can we build a simple model to describe the bounce of a gas-filled
ball? Here, we do this by starting from the principle of conservation
of energy.  The physical system we model consists of the ball and the
gas inside it.  Some energy is lost from this system (i.e., transferred
to the surroundings) when the ball bounces, since otherwise the ball
would return to its starting height.  In our model, we assume that the
energy which is not lost is stored in the compressed
gas~\footnote{Basketballs are filled with air, and tennis balls are
filled with air or nitrogen. The latter is produced during the molding
process from a mixture of sodium nitrite and ammonium
chloride~\cite{ITFballManufacture}.  The type of gas has no influence
on our model.} and in the wall of the ball, and that the energy loss
is proportional to the amount by which the ball is deformed during
impact.  The model predicts how changing the internal gas pressure
affects the bounce.

To test our model, we first experiment with a basketball.  We change
the pressure inside the basketball by pumping air in or letting air
out through the valve.  We later extend our model to describe the
behavior of a gas-filled ball as the temperature varies.  We test it
with a tennis ball heated and cooled to a range of temperatures.

Related work models how tennis balls grip and are spun up when they
strike the ground at different angles~\cite{brody, cross4}, measures
the room-temperature coefficient of restitution of tennis balls and
Superballs~\cite{cross2}, measures the temperature variation of the
coefficient of restitution of pressureless and pressurized tennis
balls between 0\textdegree C and 40\textdegree
C~\cite{RoseCoe,Downing2}, describes a testing method for tennis ball
bounce quality~\cite{brody2}, and models the impact of the coefficient
of restitution on tennis racquet collisions and serves~\cite{cross3}.

\section{Coefficient of Restitution}

\begin{figure}
    \includegraphics[width=0.47\textwidth]{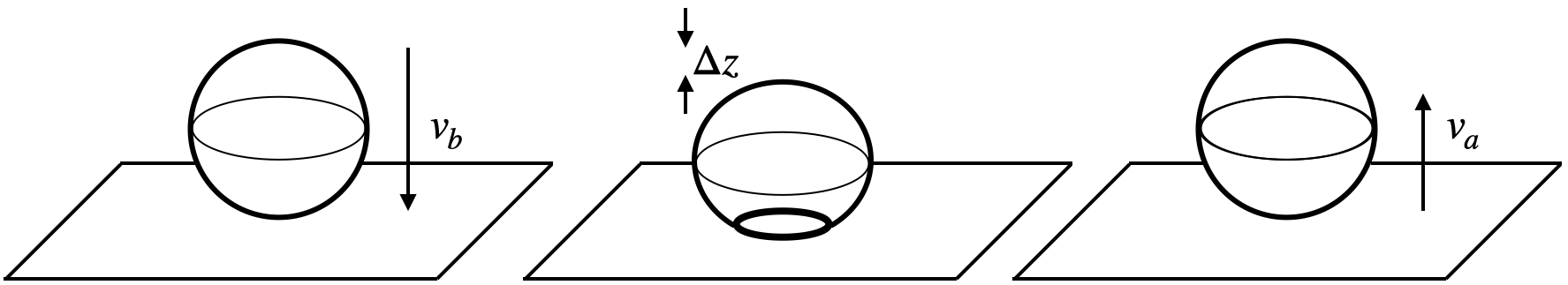}
    \caption{\label{f:bounce} The ball (a) just before the bounce with
      downwards velocity $v_b$, (b) maximally compressed by $\Delta z > 0$ in height,
      and (c) just after the bounce with (smaller) upwards velocity
      $v_a$.  The coefficient of restitution is
      $\epsilon=|v_a|/|v_b|$.}
\end{figure}

Consider a ball of mass $m$, which is released at rest from height
$h_{b}$ above a level surface and moves up and down along a vertical
line.  The subscript ``$b$'' indicates
``before the bounce''.  After release, the ball moves downwards under
the influence of gravity, accelerating until it hits the floor.
Just before that impact, denote its vertical velocity by $v_{b}$, as
shown in Fig.~\ref{f:bounce}. We neglect air resistance, so equating the
potential and kinetic energy gives $E_b = \frac{1}{2} m v_{b}^2 = m g
h_{b}$, where $g=9.8\,{\rm m/s}^2$ is the acceleration of gravity, and
$E_b$ is the energy of the ball before it hits the ground.

One way to quantify the loss of energy during the bounce is via the
``coefficient of restitution'' $\epsilon$.  This dimensionless number
is the ratio of the vertical speed of the ball just after the bounce
$|v_a|$ to the vertical speed of the ball just before the bounce
\begin{equation}
  \label{e:definee}
  \epsilon = |v_{a}|/|v_{b}|,
  \end{equation}
where subscript ``$a$'' means ``after the bounce''.  Since the speed after
the bounce is smaller than before, the coefficient of
restitution lies in the range $0 \leq \epsilon < 1$.

While it is not given this name, the coefficient of restitution is
described by Newton in
discussing the relative velocities before and after bouncing impacts
(reflexion) for balls made of wool, steel, cork, and glass.
Cook~\cite{cook} provides a detailed discussion of Newton's approach,
which is often called ``Newton’s experimental law of impacts''.  The
literature uses three symbols to denote the coefficient of
restitution: COR, $e$ and $\epsilon$.  To avoid confusion with Euler's
number $e=2.71828\cdots$ we follow the third convention.

The ratio of the ball's energy $E_a$ after the bounce to its energy
$E_b$ before the bounce can be expressed in terms of $\epsilon$.  Just
before and just after the bounce, the potential energy vanishes, and
all of the energy is kinetic, given by $\frac{1}{2} m v^2$, where $v$
is the vertical velocity. The ratio of energies is then
\begin{equation}
  \label{e:energyratio}
\frac{E_a}{E_b} = \frac{m v^2_{a}/2}{ m v^2_{b}/2} = \left( \frac{v_{a}}{v_{b}} \right)^2 = \epsilon^2,
\end{equation}
where the final equality follows from Eq.~(\ref{e:definee}).

The ratio of heights after and before the bounce may also be expressed
in terms of the coefficient of restitution $\epsilon$.  After the bounce, the ball
moves upwards, slowing down, and transforming its kinetic energy into
potential energy.  It reaches the maximum height $h_a$ at the moment
when the velocity drops to zero, so (neglecting air resistance) all of
the kinetic energy is converted to potential energy $m g h_a$.  Thus
\begin{equation}
  \label{e:heightratio}
  \frac{E_a}{E_b} = \frac{mg h_{a}}{ m g h_{b}} = \frac{h_{a}}{h_{b}}= \epsilon^2,
\end{equation}
where we have used Eq.~(\ref{e:energyratio}) to obtain the final equality.

It is hard to measure the speed of the ball just before and just after
the bounce to compute $\epsilon$ directly from the definition in
Eq.~(\ref{e:definee}). It is easier to measure the
release height and the bounce height.  So in an experiment, the
coefficient of restitution may be determined from
Eq.~{\ref{e:heightratio}) via
\begin{equation}
  \label{e:sqrt}
  \epsilon = \sqrt{\frac{h_{a}}{h_{b}}},
\end{equation}
where $h_a$ and $h_b$ are measured from the ground to the bottom of the
ball.  For a very bouncy ball, $\epsilon$ is close to 1, and for a
very ``dead'' ball, $\epsilon$ is close to 0.

\section{Modeling the energy stored during the bounce}

Our model neglects air resistance, so the energy of the ball is
constant $E_b$ before the bounce, and has a smaller constant value $E_a$ after the
bounce. The difference between these is the energy which is lost
(e.g., converted to thermal energy and sound) during the bounce.  The remaining
energy, which is stored, is $E_a$.

To model the energy stored in the bounce, first we treat the ball as a
bag containing gas under pressure.  When the ball hits the ground and
comes to a stop, the internal volume $V$ of the bag
decreases by a small amount $\Delta V \ge 0 $ to $V-\Delta V$, as shown in
Fig.~\ref{f:bounce}.  This decrease in volume requires energy, since a
force must be applied against the pressure of the gas inside the ball.
For now, assume that the material of the bag does not store or release
any energy.

\begin{figure}[htbp]
    \includegraphics[width=0.25\textwidth]{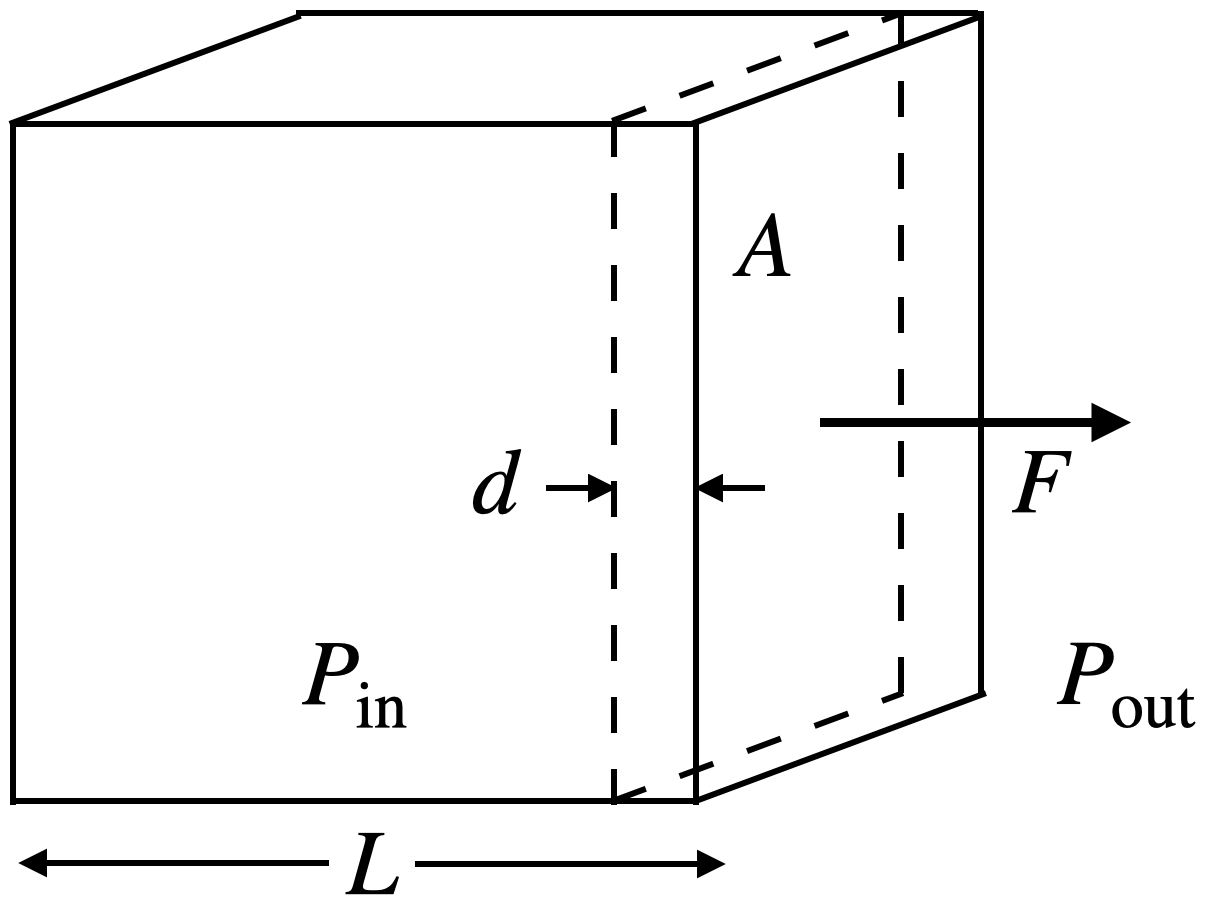}
    \caption{\label{f:box} A closed box with a sealed right wall which
      slides in and out without friction. The net force $F$ on the
      wall is the product of the area of the wall and the difference
      between the inside and outside pressures.}
\end{figure}

To compute the energy needed, consider a ``cubical box'' model of a
ball, as shown in Figure~\ref{f:box}, where one wall is free to slide
in and out, but sealed against gas loss.  This is like a bicycle pump,
in which a sliding piston compresses air, or an internal combustion
engine, where an air-fuel explosion creates pressure that moves a
piston.

Suppose that the box is filled with gas at pressure $P_{\rm in}$, and
surrounded by gas at pressure $ P_{\rm out}$.  (At sea level $ P_{\rm
  out} = 1 \, \text{atmosphere} \approx 1 \, \text{bar} = 100 \,
\text{kPa} \approx 14.7\,\text{psi}$.)  The net outwards force acting
on the moving wall is $F =P A$, where $A$ is the area of the wall, and
$P = P_{\rm in} - P_{\rm out}$ is the pressure difference.  To push
the wall in by a small distance $d$, we must do work $W = Fd = P A d$.
This reduces the volume of the box from $V=L A$ to $V-\Delta V =
(L-d)A = LA-Ad$, so the decrease in the box volume is $\Delta V = A d
$. Thus, the work we have done to change the volume (hence the
subscript ``V'') is
\begin{equation}
  \label{e:volumework}
  W_V = P \Delta V \, .
\end{equation}
This is the same for any sealed volume containing a gas, regardless of
its shape.

\begin{figure}[htbp]
    \includegraphics[width=0.4\textwidth]{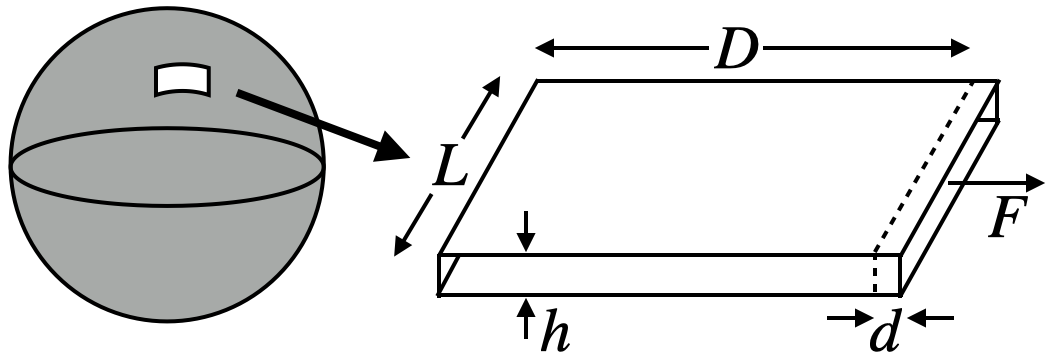}
    \caption{\label{f:sheet} A thin rectangular rubber slab of size $L
      \times D$ and thickness $h$ has internal stress $\sigma$, a
      force per unit area. To stretch it a small distance $d$ requires work
      against the force $F$ arising from the internal stress.}
\end{figure}

Now, let's look more closely at the wall of the ball.  Because the
wall is under tension from the pressure of the gas inside, increasing
its area requires energy, and decreasing its area releases energy.  To
see this, suppose that the ball wall is a thin spherical rubber shell
of thickness $h$.  Consider a bit of that material, as shown in
Fig.~\ref{f:sheet}, which is small enough to be treated as a flat
slab. In the directions tangent to the surface of the ball, this
has an internal stress (force per unit area) $\sigma$.  To stretch
the slab a small distance $d$ requires work $W = F d$.
Since $h L$ is the cross-sectional area of the slab, the force is 
$F = \sigma h L$.
Thus, the work done to change the area of the ball (hence the
subscript ``A'') is
\begin{equation}
  \label{e:areawork}
  W_A = \sigma h L d = \sigma h \Delta A \, ,
\end{equation}
where $\Delta A = L d$ is the increase in the surface area of the
ball~\footnote{It is tempting to write $W=\sigma \Delta V$, where
$\Delta V = hLd$ is the increase in the volume of the slab.  Because
rubber is almost incompressible (Poisson ratio $\nu = 1/2$) this is
incorrect. When a slab is stretched as shown in Fig.~\ref{f:sheet},
its $L$ and $h$ dimensions decrease slightly to keep the volume
unchanged.}.

For a round ball of uniform thickness, symmetry implies that the
internal stress is the same in all directions tangent to the surface,
so the work done is independent of how the area is changed.  Since the
wall of the ball is under tension, decreasing its area by an amount
$\Delta A$ releases energy $\sigma h \Delta A$.

\begin{figure}[htbp]
    \includegraphics[width=0.25\textwidth]{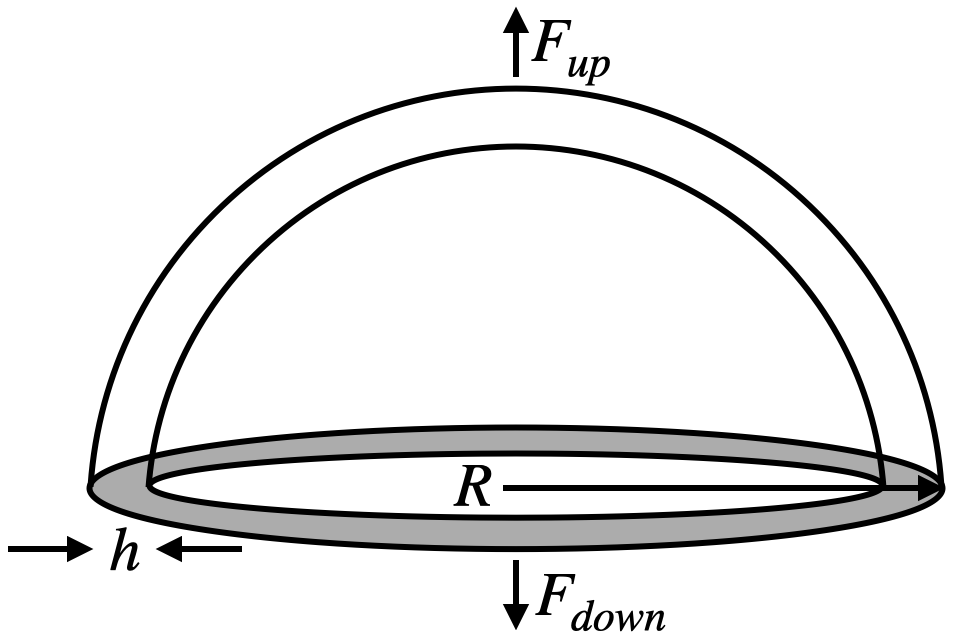}
    \caption{\label{f:HemiSphere} The difference $P = P_{\rm in} -
      P_{\rm out}$ between the internal and external pressure creates
      an upwards force $F_{up} = \pi R^2 P$ on the upper half of the
      ball wall.  That force is spread over an annular ring of area $2
      \pi R h$, shown in gray, and is canceled by a force $F_{down}$
      created by internal tension in the ball wall.}
\end{figure}

We now show that in the ball's equilibrium, the internal stress
$\sigma$ in the ball wall is proportional to the difference $P$
between the inside and outside pressure.  In equilibrium, the forces
acting on any bit of the wall must sum to zero. As shown in
Fig.~\ref{f:HemiSphere}, the pressure difference creates an upwards
force on the top half of the ball, because the horizontal components
sum to zero.  The vertical components sum to $F_{up}=\pi R^2 P$
upwards, where $R$ is the radius of the ball. This is canceled by an
equal magnitude downwards force from the bottom half of the ball,
carried by an annular ring of area $A = 2 \pi R h$. Hence, the
tangential stress $\sigma$ in the ball wall is related to the pressure
difference $P$ by $\sigma = F/A = \pi R^2 P / 2 \pi R h = (R/2h) P $.
(The ratio $R/2h$ can be very large, i.e., several thousand in an
inflated balloon.)

The energy needed to compress the ball is the difference of $W_V$ in
Eq.~(\ref{e:volumework}) and $W_A$ in Eq.~(\ref{e:areawork}).  The
first is the work done to compress the gas inside.  The second
is  energy released from the reduction in wall area. Thus,
the energy stored when the ball is compressed slightly away from its
round equilibrium is
\begin{equation}
    \label{e:stored}
  E_{a} \! = \! W_V \! - \! W_A \! = \! P \Delta V \!- \! \sigma h \Delta A \! = \! P \left( \Delta V \! - \! R \Delta A/2 \right),
\end{equation}
where the final equality uses  $\sigma  = (R/2h) P $.
Here, $P = P_{\rm in} - P_{\rm out}>0$ is the pressure
difference, $\Delta V>0$ is the decrease in the ball volume, $\Delta A>0$
is the decrease in the wall area, and
$h$ is the thickness of the (thin) wall.

Throughout this analysis, we assume that the decrease in volume
$\Delta V$ is small compared to the total volume. This means that
the internal pressure does not increase by much, so we can treat the
pressure difference $P$ as a constant in Eq.~(\ref{e:stored}).  We also assume that the decrease in the area of the wall is small
compared to the total area so we can treat the internal stress
$\sigma$ as a constant~\footnote{If $E_a = 0$ in~\eq{e:stored}, then it costs no energy
to compress or expand the ball.  That would mean that the ball is
unstable, since it could spontaneously shrink in size or expand.  At
first glance, our analysis suggests that a round uncompressed ball is
not stable if its radius is slightly reduced from $R$ to $R - \Delta
R$, while maintaining the round shape. This change decreases the
volume by $\Delta V = 4\pi (R^3 - (R - \Delta R)^3)/3 \approx 4 \pi
R^2 \Delta R$ and decreases the area (similar calculation) by $\Delta
A = 8 \pi R \Delta R$.  Thus, the two terms $\Delta V - R \Delta A/2$
which appear in Eq.~(\ref{e:stored}) cancel: in this constant-pressure
constant-tension model, energy is neither required nor released.  What
maintains the stability of the round ball are ``second-order effects''
that this model neglects: the decrease in radius slightly increases
the internal pressure, and slightly reduces the wall tension.  These
create a nonzero energy cost and thus provide stability.  Fortunately,
Eq.~(\ref{e:stored}) is sufficient to model/describe all other
possible deformations of the ball (which necessarily break spherical
symmetry).  This is because a sphere has the smallest area of
all surfaces that enclose a given volume.}.

\begin{figure}[h]
    \includegraphics[width=0.2\textwidth]{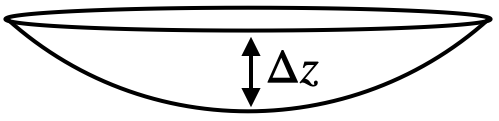}
    \caption{\label{f:bounce2} Enlarged view of the spherical cap of
      height $\Delta z>0$, from the bottom of the middle panel in
      Fig.~\ref{f:bounce}.}
\end{figure}

We now analyze the bounce shown in Fig.~\ref{f:bounce}. The bottom
part of the ball (height $\Delta z \ge 0$) is compressed so that part of
the wall becomes a section of a plane rather than a section of a
sphere, as shown in Fig.~\ref{f:bounce2}.  The reduction in ball
volume (volume of the spherical cap) is~\cite[Sec.~4.8.4]{harris}
\begin{equation}
    \label{e:dV}
    \Delta V  =   \pi R \Delta z^2 - (\pi/3) \Delta z^3.
\end{equation}
The reduction in the area of the ball wall (area of the
spherical cap minus the area of the bounding circle)
is~\cite[Sec.~4.8.4]{harris}
\begin{equation}
    \label{e:dV2}
    \Delta A  =  \pi \Delta z^2.
\end{equation}
For small $\Delta z$, consistent with our assumption that the
volume and area changes are small, the stored energy
Eq.~(\ref{e:stored}}) is
\begin{equation}
    \label{e:stored2}
    E_{a} = P(\Delta V - R \Delta A/2) \approx \pi R P \Delta z^2/2
    \approx (P/2) \Delta V,
\end{equation}
where $E_{a} \ge 0$. This shows that energy is needed to produce the
distortion shown in Fig.~\ref{f:bounce}: the energy released
by the ball wall provides only half of the amount needed to decrease
the volume of the ball.

\section{Modeling the energy lost during the bounce}

When a ball bounces, energy is lost in several ways.  The work $P
\Delta V$ done to compress the gas slightly increases the average
squared velocity of the gas molecules inside the ball.  This increases
their kinetic energy, and hence the gas temperature and pressure. That
energy is mostly recovered when the ball re-expands and the gas cools
(a ``reversible'' or ``adiabatic'' process) but a bit of the thermal energy is
lost to the environment.  Further energy is lost because the rubber
wall is deformed and ``squished'' during the bounce. This heats up the
rubber, transforming some kinetic energy into thermal energy. The
collision also creates vibrations, which are damped by internal
friction and thus converted to thermal energy.  (In tennis balls, most of the
  energy loss is due to hysteresis in the rubber material and
  intra-fiber friction in the felted shell~\cite{AshcroftStronge}.)

Here, instead of addressing the complicated details of hysteresis in
rubber~\cite{Schaefer} and other energy loss mechanisms, we make a
simplifying assumption. By analogy with Eq.~(\ref{e:stored2}), which
shows the energy stored is proportional to the change in volume
$\Delta V \ge 0$, we \emph{assume} that the energy lost when the ball
bounces is \emph{also} proportional to $\Delta V$.
(This corresponds to a damping force proportional to the velocity,
as used in~\cite{Wadhwa}. Other assumptions
are possible, for example \cite{Georgallas} assume that the energy
lost is proportional to the maximum displacement of the ball during
the bounce~\footnote{In~\cite{Georgallas} the unnumbered formulas for
$k_i$ and $k_f$ contain an $F_D x$ term; linearity in $x$ implies that
$F_D$ is a constant opposing force.}.)

With this assumption, the energy lost is
\begin{equation}
  \label{e:elost}
  E_{\rm lost} = (\mu/2) \Delta V \ge 0,
\end{equation}
where the constant of proportionality $\mu/2 \ge 0 $ is the energy
lost per unit deformation of the volume.  We refer to $\mu$ as the
\emph{loss coefficient}.  The factor of one-half is for convenience:
it simplifies the form of the equations that follow.  Bouncy balls,
which return to almost the drop height, have a small $\mu$, so very
little energy is converted to thermal energy.  Balls with a large
$\mu$ generate a lot of thermal energy for a small change in volume,
resulting in a weak bounce.  The model of Eq.~(\ref{e:elost}) includes
all energy loss mechanisms: each one contributes to $\mu$.

\section{Pressure dependence of the coefficient of restitution}

We can now determine the pressure dependence of the bounce.
Conservation of energy implies that
\begin{equation}
\label{e:alt}
  E_{b} = E_a + E_{\rm lost} = (P/2) \Delta V +  (\mu/2) \Delta V,
\end{equation}
where we have used Eqs.~(\ref{e:stored2}) and (\ref{e:elost}).  If we
divide Eq.~(\ref{e:stored2}) by Eq.(\ref{e:alt}), $\Delta V$
cancels, yielding a simple relationship for the coefficient of
restitution
\begin{equation}
\label{e:almostfinal}
\epsilon^2 = \frac{E_a}{E_b}  = \frac{P_{\rm in} - P_{\rm out}}{P_{\rm in} - P_{\rm out} + \mu},
\end{equation}
where the first equality follows from Eq.~(\ref{e:energyratio}) and
the pressure difference $P = P_{\rm in} - P_{\rm out}$.

A nice feature of this model: it predicts that the coefficient
of restitution $\epsilon$ is independent of the ball's velocity.  This
is typically the case if the velocities are not too
large~\cite[Fig.~8]{Bridge1}.  Such smaller velocities are also
consistent with our assumption that $\Delta V$ (the change in ball
volume during the bounce) is small.
 

\section{Experimental testing and revised model}

To test our model, we drop a basketball from the top of a door ($h_b=2.05$m)
and video the bounce with a mobile phone.  This is repeated three
times for 15 different internal pressures. We measure the images
with a vernier caliper, scale them to determine the bounce height,
then infer $\epsilon$ using Eq.~(\ref{e:sqrt}) and estimate its
uncertainty $\Delta \epsilon$.  We then compare to the model.



The basketball pressure is measured using a dial-type pressure gauge
reading in pounds per square inch (psi) which displays the difference
$P_{\rm in} -P_{\rm out}$ between the pressure inside and outside the
ball.  The results of the bounce tests are shown in
Table~\ref{t:dataBasketball}, for pressure differences up to 13~psi,
taking $P_{\rm out} = 14.7\, \text{psi}$.  Out of fear that it might
explode, we did not inflate the basketball to higher
pressures~\footnote{{\bf Warning:} bursting a basketball via
overpressure is dangerous and can lead to serious injuries such as
permanent loss of sight or hearing.}.

As can be seen from the first row of Table~\ref{t:dataBasketball},
even with the ball deflated (inside and outside pressures equal)
it still bounces weakly (34cm, 17\% of its starting height).  In
contrast, our model incorrectly predicts that if $P_{\rm in} = P_{\rm out}$,
then there is no bounce, since Eq.~(\ref{e:almostfinal}) gives $\epsilon =
0$.

Clearly, our model needs correction, so we revisit its assumptions.
Videos show that the deflated ball does not deform much when it
bounces, so our assumption that the change in volume is small is
satisfied. We conclude that the thin rubber shell of the ball is
storing energy, and hence adjust our model, treating this stored
energy as if it were due to an additional fictitious gas pressure
$\mu_1$ inside the ball. If this is a faithful model, then
Eq.~(\ref{e:almostfinal}) is still valid, but with an additive offset
$\mu_1$ to the internal pressure: $P_{\rm in} \rightarrow P_{\rm in} +
\mu_1$.  Thus, the revised model takes the form
\begin{equation}
\label{e:almostfinal2}
\epsilon =  \sqrt{\frac{P_{\rm in} - P_{\rm out} + \mu_1 }{P_{\rm in} - P_{\rm out} + \mu_2}},
\end{equation}
where $\mu_2 = \mu + \mu_1$.  The constant $\mu_1$ characterizes the
energy storage in the rubber, whereas $\mu_2$ is determined by both
the energy storage in the rubber and by its loss coefficient. 

\begin{table}
 \begin{tabular}{c|c|c|c} 
Index i & ~  $P_{\rm in}$ (psi) ~ & ~ ~ ~ $\epsilon$ ~ ~ ~ & ~ ~ $\Delta \epsilon$ ~ ~ \\
   \hline
1  & 14.7   &  0.407  &  0.009 \\
2  &15.7   &  0.582  &  0.012 \\
3  &16.7   &  0.674  &  0.014 \\
4  &17.7   &  0.718  &  0.015 \\
5  &18.45  &  0.727  &  0.015 \\
6  &18.7   &  0.755  &  0.016 \\
7  &19.7   &  0.768  &  0.016 \\
8  &20.7   &  0.783  &  0.016 \\
9  &21.7   &  0.806  &  0.016 \\
10 &22.7   &  0.824  &  0.017 \\
11 &23.7   &  0.836  &  0.017 \\
12 &24.45  &  0.834  &  0.019 \\
13 &25.7   &  0.856  &  0.017 \\
14 &26.45  &  0.865  &  0.017 \\
15 &27.7   &  0.869  &  0.018 \\
    \hline
 \end{tabular}
   \caption{\label{t:dataBasketball} The coefficient of restitution
     $\epsilon$ and its uncertainty $\Delta \epsilon$ for a basketball
     at 15 different internal pressures.}
\end{table}

To test the model, we fit the data of Table~\ref{t:dataBasketball} to
Eq.~(\ref{e:almostfinal2}).  We find the values of the unknown
constants $\mu_1$ and $\mu_2$ with a standard $\chi^2$
statistic, which quantifies the deviations between the model and the
data.  $\chi^2$ is the sum of the squared differences between the
model and data, weighted by the uncertainty in the measurements,
\begin{equation}
  \label{e:chi2def}
  \chi^2 = \sum_{i=1}^{15} \Bigl(\,  \epsilon(P_{{\rm in},i}) - \epsilon_i  \, \Bigr)^2/\Delta \epsilon^2_i\, .
\end{equation}
The sum is over the 15 data points in Table~\ref{t:dataBasketball} and
the function $\epsilon(P_{\rm in})$ is given by
Eq.~(\ref{e:almostfinal2}).  The best-fit values of $\mu_1$ and
$\mu_2$ are the ones which minimize $\chi^2$.

\begin{figure}
  \includegraphics[width=0.5\textwidth]{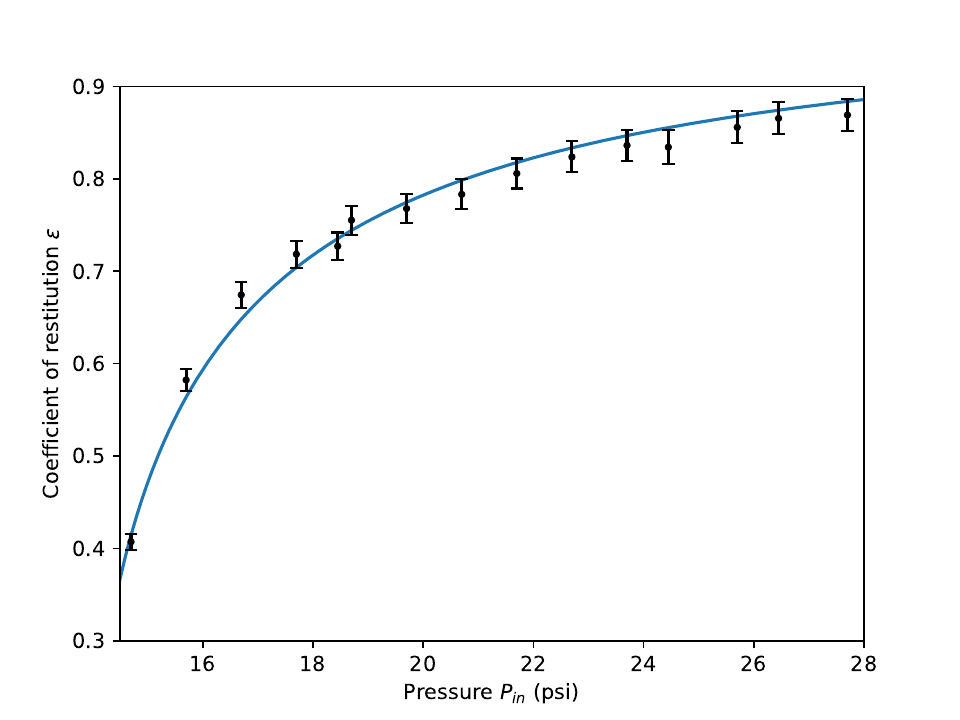}
  \caption{\label{f:plotBB} The basketball data of
Table~\ref{t:dataBasketball} compared with the 
    best-fit model of Eq.~(\ref{e:almostfinal2}).}
\end{figure}

To find $\mu_1$ and $\mu_2$, we wrote a short Python notebook which
computes $\chi^2$ on a dense rectangular grid of $\mu_1$ and $\mu_2$
values.  The minimum of $\chi^2 = 13.4$ is found at $\mu_1 = 0.80
\text{ psi}$ and $\mu_2 = 4.67 \text{ psi}$.  The data and the
best-fit model are plotted in Fig.~\ref{f:plotBB} and are in good
agreement.
Since $\mu_1$ is smaller than typical values of $P_{\rm in} - P_{\rm
  out}$, the bounce arising from the fictitious gas pressure is small.

If the model were exact and the experimental errors were distributed
as independent Gaussian random variables, then with 90\% confidence a
$\chi^2$ value (13 degrees of freedom) of less than $19.8$ would be
obtained.  So, the $\chi^2$ value of $13.4$ indicates a good fit,
consistent with the estimated uncertainties.

\section{Temperature dependence of the coefficient of restitution}

We were curious if our model also describes the temperature-dependence
of the coefficient of restitution for a sealed ball.  In this case,
the ideal gas law implies that the pressure of a fixed number of gas
molecules (those inside the ball) is proportional to the absolute gas
temperature.  Thus, changing the temperature of the gas inside the
ball changes $P_{\rm in}$.

If we assume that the loss coefficient $\mu$ does not change with
temperature, then we can show how $\epsilon$ depends upon the
temperature $T$ (in Kelvin) of the gas in the ball.  From the
ideal gas law, $P_{\rm in} = k T$. Here, $k$ is a constant: the
product of Boltzmann's constant and the number of gas molecules per
unit volume inside the ball.  Substitute this into
Eq.~(\ref{e:almostfinal2}), removing the common factor of $k$ from
numerator and denominator.  One obtains the dependence of $\epsilon$
on temperature $T$:
\begin{equation}
\label{e:final}
\epsilon(T) = \sqrt{\frac{T-{\mathcal{T}}_1}{T-{\mathcal{T}}_2}}.
\end{equation}
Here ${\mathcal{T}}_1 = P_{\rm out}/k$ and ${\mathcal{T}}_2 = (P_{\rm
  out}-\mu)/k$ are constants with units of Kelvin.  These constants
depend upon the outside pressure, $k$, and $\mu$.  Although we
don't know these values, we can measure the coefficient of restitution
$\epsilon$ experimentally at several different temperatures and fit
Eq.~(\ref{e:final}) to the data to obtain the values of
${\mathcal{T}}_1$ and ${\mathcal{T}}_2$.

To test Eq.~(\ref{e:final}), we drop a tennis ball at different
temperatures from the top of a door ($h_b=2.05$m) and video the bounce
with a mobile phone.  This is repeated three times for each of seven
different temperatures spanning a 102K range from -22\textdegree
C to 80\textdegree C.  A storage freezer, a refrigerator and an oven
are used to cool and heat the ball. The data are shown in
Table~\ref{t:data}.

\begin{table}[h]
 \begin{tabular}{c|c|c|c} 
~Index i & ~  $T$ (Kelvin) ~ & ~ ~ ~ $\epsilon$ ~ ~ ~  & ~ ~ $\Delta \epsilon$ ~ ~ \\
 \hline
 1 & 251  & 0.447  & ~0.023 \\ 
 2 & 274  & 0.621  & ~0.014 \\
 3 & 278  & 0.629  & ~0.005 \\
 4 & 293  & 0.704  & ~0.009 \\
 5 & 323  & 0.766  & ~0.008 \\
 6 & 343  & 0.801  & ~0.021 \\
 7 & 353  & 0.813  & ~0.004 \\
 \hline
 \end{tabular}
   \caption{\label{t:data} The coefficient of restitution $\epsilon$ for a
     tennis ball at different temperatures, and the estimated
     uncertainty $\Delta \epsilon$. }
\end{table} 

After those trials were complete, we punctured the tennis ball
(drilling a 4mm diameter hole) to release all of the internal
pressure.  At room temperature the ball still bounced, showing that
(as for the basketball) the rubber body stores some energy.  We again
model this as an offset to the internal pressure.  This does not
change the form of Eq.~(\ref{e:final}), but modifies the
interpretation of the constants $\mathcal{T}_1 = (P_{\rm out} -
\mu_1)/k$ and $\mathcal{T}_2 = (P_{\rm out} - \mu_2)/k$.

Some manufacturers produce ``pressureless'' tennis
balls~\cite{PressurelessBall}, which get their bounce from a spherical
rubber core.  Downing~\cite{Downing} and Rose and Coe~\cite{RoseCoe}
show that their $\epsilon$ behaves differently than that of
pressurized balls, so our model is unlikely to apply. One use of
pressureless balls is at high altitudes, where normal tennis balls
bounce higher than at sea level.  This follows from
\eq{e:almostfinal2}: increasing $P_{\rm in} - P_{\rm out}$ increases
$\epsilon$.  Indeed, there are ``high altitude'' pressurized balls,
which correct for this by having smaller $P_{\rm in}$ than normal
balls.

\begin{figure}
  \includegraphics[width=0.5\textwidth]{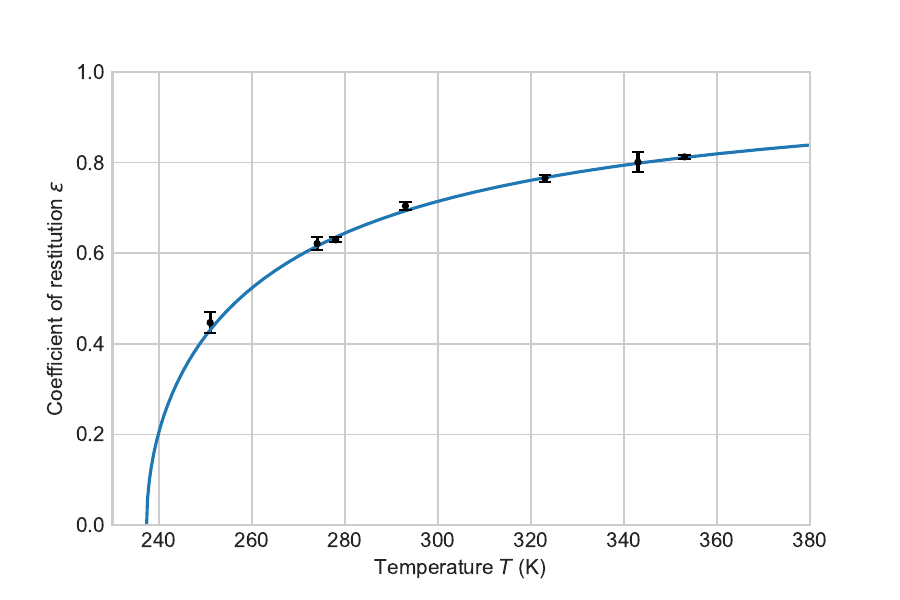}
  \caption{\label{f:plot} The tennis ball data of Table~\ref{t:data}
    and best-fit model of Eq.~(\ref{e:final}).}
\end{figure}

Again, the model of Eq.~(\ref{e:final}) is fit to the data of
Table~\ref{t:data} using a $\chi^2$ statistic.
The sum in Eq.~(\ref{e:chi2def}) is now over the seven data
points in Table~\ref{t:data}, and the modeled $\epsilon$ is the
function of temperature $T_i$ given in Eq.~(\ref{e:final}).  The
minimum value of $\chi^2$ is at ${\mathcal{T}}_1=237.3$K and
${\mathcal{T}}_2=177.3$K, for which $\chi^2 = 3.3$.

The data and the best-fit model are plotted in Fig.~\ref{f:plot} and
are in good agreement.  If the model were exact and the experimental
errors were distributed as independent Gaussian random variables, then
with 90\% confidence a $\chi^2$ value (five degrees of freedom) of
less than $9.2$ would be obtained.

Our experiment used an old, ``dead'' tennis ball.  A fresh ball, which
will have a higher internal pressure, should have a smaller value of
${\mathcal{T}}_1$.


Since the model and data agree well, one might think that the
assumptions used to model the temperature dependence of tennis balls
are correct.  Previous studies of tennis balls indicate that this is
not the case.
For example, we assume that the rubber properties, such as the
loss coefficient $\mu$, are temperature independent.
Consequently, our model predicts a decreasing contact time as
the temperature (and hence internal pressure) increase.  However,
more detailed studies of tennis
balls~\footnote{See~\cite{Downing2,ThomasBruceAllen}, but note that
this data is mainly collected for the larger ball velocities and
smaller temperature range found in the sport. For example, in our
experiment, the ball velocity at the ground is a bit less than
$6\,\text{m/s}$; in competitive tennis, it can reach
$65\,\text{m/s}$.} show that the properties of the rubber change
significantly and that the contact time
increases~\cite{Downing,Downing2} with temperature over the range
10-40\textdegree C.  Indeed, the International Tennis Federation
states that ``high temperatures can affect the ball rubber, thus
increasing bound height''~\cite{ITFballApprovalTests}.


Since some of its assumptions are not satisfied, it is remarkable that
the model fits the temperature data so accurately.  This may be
because the model's functional form applies more broadly. For example,
we have assumed that the gas stores the energy during the bounce.
However, the same functional form is obtained regardless of where
energy is stored, provided that the energy stored and the energy lost
are proportional to the change in volume $\Delta V$ and are linear
functions of the pressure and/or temperature.


\section{Comparison with other data/models}

The published literature contains data showing how the coefficient of
restitution $\epsilon$ varies with pressure for several types of thin
gas-filled balls, and some analytic and numerical models.  These
include data for play balls~\cite{Bridge1,Bridge2}, soccer
balls~\cite{Georgallas}, basketballs~\cite{Georgallas}, and
volleyballs~\cite{Georgallas}. We checked our model against these.

Our model is a good fit to play ball data taken by
Bridge~\cite{Bridge1}.  This shows the ``bounce efficiency''
$\epsilon^2$ for pressure differences $P=P_{\rm in} - P_{\rm out}$
ranging from $3$ to $146 \, \text{kPa}$, with the ball dropped from
$1\, \text{m}$ height.  Bridge fits the data \cite[Fig.~9]{Bridge1} to
a two-parameter $(P_0,n)$ power-law model $\epsilon^2(P) = 1/(1 +
(P/P_0)^n)$; the best fit has $n=-0.62$. However, this model is
ad-hoc, and not derived from physical principles.  Our model fits the
data about as well as the power law model, although the lack of error
bars in Fig.~9 of Bridge~\cite{Bridge1} makes it impossible to
quantify if the fit of our model is consistent with the
observational/experimental uncertainties.  The same data is shown in
Fig.~6 of a later paper by Bridge~\cite{Bridge2}, where it is compared
to a cylindrically-symmetric numerical model.  Our model fits the data
somewhat better than that numerical model.

We also compared our model to basketball, soccer ball and volleyball
data taken by Georgallas and Landry (G\&L) for drop heights of 0.75m
and 1.5m~\cite[Fig.~3]{Georgallas}. Note that while G\&L derive a
model for $\epsilon$, and test it against their data, they exclude
data points with ``gauge pressure'' $P=P_{\rm in} - P_{\rm out}=0$
from their fits.  Indeed, the lowest-pressure points deviate
significantly from their model predictions. G\&L argue that this is
expected, since the balls are significantly deformed at low
pressure. Their model, like ours, assumes small deformations.

\emph{Basketball}: The G\&L fits are good, apart from three data
points with $P < 10\,$kPa; the bottom two points differ from the model
by many standard deviations.  Our model is a comparable fit for $P >
10\,$kPa and is a better fit at smaller $P$, either including or
excluding the $P=0$ data point from the fit.

\emph{Soccer ball}: The G\&L fits are good for the larger $P$ values
but are very poor for the three points with $P < 10\,$kPa; we suspect
that these were ignored or excluded by their fitting procedure.  If we
also exclude these points, then our model fits about as well as their
model.

\emph{Volleyball}: The G\&L fits are poor for the two points with $P <
6\,$kPa but good for higher pressures. Even when we drop these points
(as G\&L appear to have done) our model does not fit the data quite as
well as theirs. Our model predicts $\epsilon$ values that are systematically
about half of a standard deviation too low for $20 < P/\text{kPa} <60$
and that are about the same amount too high for $80 < P/\text{kPa}$.


It is interesting to compare the G\&L analytic model~\cite{Georgallas}
to ours. Other work analyzes forces, see \cite{hubbard,AshcroftStronge2} and references therein. 
G\&L's is the only model we have found which is also derived
using conservation of energy. The key difference is the assumed model
for energy loss.  G\&L model this as due to a ``constant
dissipative force'' $F_D$, whose direction is opposite to the ball
velocity.  Thus, the energy loss $2 F_D \Delta z$ is
proportional to the deformation $\Delta z$ at maximum compression
(note: G\&L denote $\Delta z$ by $x$ and $\epsilon$ by $e$).  This energy loss is
proportional to $\sqrt{\Delta V}$, since Eq.~(\ref{e:dV}) shows that
the reduction of the ball volume at maximum compression $\Delta V
\propto \Delta z^2$ for small $\Delta z$.  In contrast, our model
assumes that the energy loss is proportional to $\Delta V$.  If G\&L
modeled energy loss as we do, then the denominator
of~\cite[Eq.~(6)]{Georgallas} becomes $1-e^2$ rather than
$(1-e^2)^2$, and the G\&L model reduces to ours.

The G\&L model has a troubling feature. All but one of the
``constants'' that appear in their energy transformation equations
($m$, $g$, $R$, $G$, $D_W$, $F_D$) depend upon the ball shape and
material properties: they are indeed constants.  The exception is
$F_D$, which is \emph{not} a constant: It has an unspecified velocity
dependence. This means that their model is incomplete.  Unlike our
model, which predicts the same $\epsilon$ for \emph{all} velocities,
their model contains a parameter $F_D$ which must be determined
experimentally for each different drop height/incoming
velocity~\cite[Table 2]{Georgallas}.  The model is not predictive
because it does not specify how $F_D$ varies with drop height or
incoming velocity.  For example, we can not determine if their model
predicts a velocity-independent $\epsilon$ for small velocities.

The literature also contains other models for the energy loss.  One is
based on ``momentum flux'' in the ball.  This gives rise to a
dissipative force proportional to the square of the instantaneous
velocity during the bounce. The force only acts when the ball is
moving upwards \cite[the $\dot z^2 H(-\dot z)$ term in
  Eq.~11]{hubbard} giving rise to an energy loss proportional to
$\Delta V^{3/2}$.  In contrast, the G\&L model corresponds to a force
of constant magnitude oriented opposite to the instantaneous velocity
leading to an energy loss proportional to $\Delta V^{1/2}$. Our model
corresponds to a dissipative force, acting in both directions,
proportional to the instantaneous velocity, leading to an energy loss
proportional to $\Delta V$. The resulting equations of motion are
those of a standard damped harmonic oscillator (dissipative force
linear in velocity).  This model was also adopted in~\cite{Wadhwa},
and used to characterize tennis balls, superballs, soccer balls,
squash balls and table-tennis balls.
 


\section{Conclusion}

We have derived a simple two-parameter model which describes how the
coefficient of restitution $\epsilon$ of a gas-filled ball varies with
pressure.  Our model, which we have not found in the literature, is a
good fit to data from many types of gas-filled balls.  We expect that
it will work well for any type of thin-wall gas-filled ball, provided
that the volume deformations are small.

A fundamental assumption of our model is that the energy loss is
proportional to the maximum change in the volume of the ball during a
bounce. We expect that this would break down when the height
is large enough that the ball is significantly deformed.  Note that
our test velocity (around 6~m/s) is an order of magnitude smaller than
the velocity reached during many competitive sports. For example, see
Fig.~3.9 of~\cite{ThomasBruceAllen} which shows the maximum
deformation of a tennis ball expected at 30~m/s.  The assumptions of
our model could be explored using such finite element simulation
models~\cite{AllenGoodwillHaake,ThomasBruceAllen}.

When we extended our model to describe how $\epsilon$ changes as a
function of temperature, we assumed that the loss coefficient $\mu$ is
a temperature independent constant. While this is incorrect, the model
agrees surprisingly well with the data. Indeed,
Feynman remarks that simple models sometimes work better and apply
more broadly than expected.  In that spirit, it would be interesting
to further examine our model. For example, to see if the model (which
was developed for gas-filled balls) correctly describes the behavior
of other types of balls such as solid rubber Superballs. We anticipate
that it won't work well, because these balls store most energy in the
compression of their rubber bodies rather than in the compression of
gas. Then, the behavior of the rubber would govern the balance between
energy stored and energy lost.

\def\myspace{\vspace{-0.03 in}}

\begin{acknowledgments}

\myspace
We thank A. Jones for his encouragement and support during the IB
program, L. Nieder, M.A. Papa and J. Romano for helpful comments on a
draft of this manuscript, and the anonymous referees for pointing out
some significant errors in an earlier version of this manuscript.
\end{acknowledgments}

\myspace
\section*{Conflict of interest statement}
\myspace
The authors have no conflicts to disclose.
\myspace
 
\bibliography{references}
\end{document}